# Generation of rectangular optical waves by relativistic clipping

**Sándor Varró**
Wigner Research Centre for Physics, Hungarian Academy of Sciences,
Institute for Solid State Physics and Optics, Budapest, Hungary
E-mail: varro.sandor@wigner.mta.hu

**Abstract.** Theoretical results are reported, concerning the reflection and transmisson of few–cycle laser pulses on a very thin conducting layer, which may represent the surface current density of the massless relativistic charges of graphene. It is shown that the pulse may undergo violent distortions to that extent, that the scattered radiation contains rectangular trains, which are approximate physical realizations of Rademacher functions in the optical or terahetz regime.

## 1. Introduction

Laser technology has been undergoing a very fast development recently, making it possibile to generate extreme (ultrashort, very high-intensity) and well-controlled electromagnetic fields in the optical regime. Carrier-envelope phase difference effects (Milošević *et al*, 2006) became important in both diagnosing and controlling such short, e.g. attosecond pulses (Krausz and Ivanov, 2009). After the real-time observation of electron tunnelling from atoms (Uiberacker *et al*, 2007), even the tiny delays of photoemission signals, stemming from different atomic states has been measured (Schultze *et al*, 2010), and further studies of the detailed dynamics of photoemission (Nagele *et al*, 2011) is going on. Important experimental results on high-order harmonic generation in the bulk crystals (Ghimire *et al*, 2011), and attosecond time-resolved photoemission from solid targets (Neppl *et al*, 2012) raise questions in the theoretical description of non–linear electromagnetic responses of condensed matter (see e.g. Zhang and Thumm (2011), Korbman *et al* (2012), Földi and Benedict (2012)). Recently Wirth *et al* (2011) succeeded synthesizing light transients for producing sub–cycle pulses, with unprecedented control of the amplitudes and phases, which open a wide range of possibilities for studying very fast processes in atomic and condensed matter systems.

The generation of broad-band radiation and short pulses relies on the highly nonlinear processes induced by the intense laser fields. In general, the magnitudes of these nonlinearities (Fedorov, 1997) depend also on the target, of course. It is clear that –depending on various parameters– intense fields may cause relatively modest effects, and moderately intense lasers may induce very high-order processes. Here we discuss an example for the latter situation, and show that the *collective radiation back-reaction* of relativistic convective surface currents driven by a laser field can cause a violent distortion in the scattered radiation. Similar phenomena have already been considered by us, for desribing the reflection of laser pulses on thin conducting nano-layers (Varró, 2004, 2007a-b-c and 2012). The effectively massless charge carriers, electrons and holes in graphene (see Novoselov *et al* (2005), Bostwick *et al* (2007), Castro Neto *et al* (2009)), seem to be also good candidates for illustrating the high 'susceptibility' we just mentioned. The unique optical properties of graphene (Blake *et al* (2007)) are remarkable, also in the sense that they are directly related to the fine structure constant (Nair *et al*, 2008), which naturally appears in linear response theory (see e.g. Abergel and Fal'ko (2007), Cserti and Dávid (2010)). In the present paper we report on our theoretical results concerning the reflection and transmisson of a few-cycle laser pulse on a very thin conducting layer, and illustrate the temporal evolution of the electric field strength of the reflected radiation in the nonlinear regime. By working out a classical desription, we show that the pulses may undergo violent distortions, to that extent that the scattered radiation contains rectangular trains, being physical realizations of Rademacher functions in the optical or terahertz regime.

In Section 2 we derive the gauge–invariant equation of motion of massless charges. We have felt it instructive to include a derivation in a more general frame, than is in fact will be needed for the special problem we solve. In Section 3 the matching equations for the electromagnetic fields, and the force terms (including the radiation reaction term) in the equation of motion of the massless charges will be considered. In Section 4 the nature of the 'relativistic clipping effect' will be discussed, and some illustrative numerical examples and figures will be presented. In Section 5 a brief summary including conclusions closes our paper.

## 2. The gauge–invariant equation of motion of massless charges

In order to write down the relativistic equation of motion of a charged particle with zero rest mass, one needs additional considerations beyond the description of 'ordinary' massive electrons. We would like to illustrate how can one interrelate the velocity and the momentum of a massless particle, in the frame of the Lagrangian description. The results of this section will be used in describing the radiative coupling and back–reaction of the massless charges of a thin layer.

Let us consider the large scale motion of a charged particle moving in a potential $U$. with a position-dependent mass $m_0 = m_0(\mathbf{r})$ The Lagrange function $L(\mathbf{r}, \mathbf{v}, t)$ of the particle in the additional (e.g. radiation) fields $\mathbf{E} = -\nabla A_0 - \partial \mathbf{A} / \partial ct$ and $\mathbf{B} = \nabla \times \mathbf{A}$ then reads

$$L(\mathbf{r}, \mathbf{v}, t) = -m_0(\mathbf{r}) \mathbf{v}_F^2 \sqrt{1 - v^2 / v_F^2} + \frac{e}{c} \mathbf{v} \cdot \mathbf{A}(\mathbf{r}, t) - eA_0(\mathbf{r}, t) - U, \quad \mathbf{p} \equiv \frac{\partial L}{\partial \mathbf{v}} = \frac{m_0(\mathbf{r}) \mathbf{v}}{\sqrt{1 - v^2 / v_F^2}} + \frac{e}{c} \mathbf{A}(\mathbf{r}, t), \quad (1)$$

where $e$ is the charge of the particle, $c$ the velocity of light in vacuum, and $v_F$ is a characteristic, maximum velocity of the particle (for graphene $v_F = c/300$ approximately). $A_0$ and $\mathbf{A}$ are the scalar and vector potentials, respectively, which represent the incoming and the scattered radiation. In the second equation of (1) we have defined the *canonical momentum* $\mathbf{p}$. The *kinetic momentum* $\boldsymbol{\pi}$, is introduced, too, because it is a *gauge–invariant* quantity (in contrast to $\mathbf{p}$). After straightforward calculations, we receive the following kinematic relations, which will be used later on,

$$\boldsymbol{\pi} \equiv \mathbf{p} - \frac{e}{c} \mathbf{A}(\mathbf{r}, t) = \frac{m_0 \mathbf{v}}{\sqrt{1 - v^2 / v_F^2}}, \quad \frac{|m_0| v_F}{\sqrt{1 - v^2 / v_F^2}} = \sqrt{\boldsymbol{\pi}^2 + m_0^2 v_F^2}, \quad |m_0| v_F \frac{\boldsymbol{\pi}}{\sqrt{\boldsymbol{\pi}^2 + m_0^2 v_F^2}} = m_0 \mathbf{v},$$

$$\mathbf{v} \cdot \frac{\partial L}{\partial \mathbf{v}} - L = \frac{m_0 v_F^2}{\sqrt{1 - v^2 / v_F^2}} + eA_0(\mathbf{r}, t) + U. \quad (2)$$

From the third equation of (2) one can immediately see that in the limit of zero rest mass the velocity modulus reaches its maximum value and the particle has an ultrarelativistic kinematics (i.e. the energy is linearly proportional with the momentum). By taking the limits $m_0 \to \pm 0$ (and assuming that $\boldsymbol{\pi} / |\boldsymbol{\pi}|$ is unambiguosly definable some way, even for $\boldsymbol{\pi} = 0$), we have

$$\lim_{m_0 \to \pm 0} \mathbf{v} = \pm v_F \frac{\boldsymbol{\pi}}{|\boldsymbol{\pi}|}, \quad \lim_{m_0 \to \pm 0} v = v_F. \quad (3)$$

Thus, close to the points where $m_0 \to \pm 0$, we are allowed to represent the velocity by the gauge–invariant expression $\mathbf{v} = \pm v_F \boldsymbol{\pi} / |\boldsymbol{\pi}|$. The Hamiltonian $H = \mathbf{v} \cdot \partial L / \partial \mathbf{v} - L$ of the particle can be brought to the form

$$H(\mathbf{r}, \mathbf{p}, t) = (m_0 / |m_0|) v_F \sqrt{(\mathbf{p} - e\mathbf{A}(\mathbf{r}, t)/c)^2 + m_0^2(\mathbf{r}) v_F^2} + eA_0(\mathbf{r}, t) + U, \qquad \frac{d\mathbf{r}}{dt} = \frac{\partial H}{\partial \mathbf{p}}, \quad \frac{d\mathbf{p}}{dt} = -\frac{\partial H}{\partial \mathbf{r}}, \qquad (4)$$

where the last two equations are the usual canonical equations. On the basis of (4) we derive

$$\frac{d}{dt} \frac{m_0(\mathbf{r})\mathbf{v}}{\sqrt{1 - v^2/v_F^2}} = \frac{d\boldsymbol{\pi}}{dt} = \mathbf{F} + \frac{e}{c} \mathbf{v} \times \mathbf{B} - \{\nabla m_0(\mathbf{r})\} v_F^2 \sqrt{1 - v^2/v_F^2}, \qquad \mathbf{F} \equiv e\mathbf{E} - \nabla U. \qquad (5)$$

Using the fourth equation in (2) along an actual trajectory, one can introduce an *effective mass* $\tilde{m}$, by subtracting the potential energy from the total energy,

$$\frac{m_0}{\sqrt{1 - v^2/v_F^2}} = \frac{E_k}{v_F^2} \equiv \tilde{m}, \qquad E_k \equiv H - eA_0 - U, \qquad \lim_{m_0 = 0} \frac{m_0}{\sqrt{1 - v^2/v_F^2}} = \tilde{m}_0. \qquad (6)$$

Equation (6) shows that if $E_k$ is non-zero and bounded, the effective mass stays finite, even if $v \to v_F$. If $H - eA_0$ does not explicitely depend on time, then the dynamical equation (5) can be brought to a Newtonian form, by using the definition of the *effective mass* $\tilde{m}$ and the *acceleration* $\mathbf{a} = d\mathbf{v}/dt$,

$$\tilde{m} \frac{d\mathbf{v}}{dt} = (1 - v^2/v_F^2)\mathbf{F} + \frac{e}{c} \mathbf{v} \times \mathbf{B}' - \{\nabla m_0(\mathbf{r})\} \sqrt{1 - v^2/v_F^2}, \qquad \mathbf{B}' \equiv \mathbf{B} - (1/c)\mathbf{v} \times [(c/v_F)^2(\mathbf{E} - \nabla U/e)]. \qquad (7)$$

We note that the second force term in (7) also contains an electric precession term $\propto \mathbf{v} \times \mathbf{E}$, and a coupling term of the form $\propto \mathbf{v} \times [\mathbf{r} \times \mathbf{v}] dU/rdr$, which is proportional to the angular momentum (for a central potential $U$). It is interesting to note, that in case of a planar motion, according to (3), if $m_0 \to 0$ at an 'edge' along a line, say $y = y(x)$ in the $x - y$ plane, then the only term $\propto \mathbf{v} \times \mathbf{B}'$ survives, because $v \to v_F$ in this case. Thus, there is no acceleration along the line, and this latter term is always perpendicular to the velocity of the 'incoming particle'. This means that there is *no forward and backward scattering* of the particles, thus, they are forced to move along the line $y = y(x)$. This may be a simple classical picture to the so–called 'edge states', which have recently received a great importance in the context of 'topological insulators' (see Hasan and Kane (2010), Asbóth (2012)). We also note that recently there has been several investigations appeared discussing the equation of motion of quantum particles with position dependent mass (see, e.g. the paper by Cruz y Cruz (2008) on a one-dimensional non-relativistic oscillator with position dependent mass).

In the next section the equation of motion of the active charges in the $x - y$-plane of a thin layer shall be taken as the following special form of equation (5)

$$\frac{d\boldsymbol{\pi}}{dt} = e\mathbf{E} + \frac{e}{c} \mathbf{v} \times \mathbf{B}, \qquad \mathbf{v} = +v_F \frac{\boldsymbol{\pi}}{|\boldsymbol{\pi}|}. \qquad (8)$$

## 3. Matching equations for the electromagnetic fields, and the equation of motion of the massless charges in a thin layer, interacting with them

In the present section we derive and solve the matching equations for the electric field and the magnetic induction representing the scattered radiation. First we shall express the reflected and transmitted fields in terms of an unknown surface current, and then we put the total field to the equation of motion of the massless charge elements, of which the surface current consists. In this way, the sytem of equations becomes closed, and the problem is reduced to the soulution of the equation of motion for the charges.

This approach automatically incorporates a 'collective radiative back–reaction' of the complete layer, and delivers a generalization of the Fresnel formulae (Varró 2004, 2007a-b-c and 2012).

The components of a TM (p-polarized) configuration of waves $(0, E_y, E_z)$ and $(B_x, 0, 0)$, in the two separated media (in regions 1 and 3), satisfy the following Maxwell equations,

$$\partial_z B_x = \partial_0 \varepsilon \cdot E_y, \quad -\partial_y B_x = \partial_0 \varepsilon \cdot E_z, \quad \partial_y E_z - \partial_z E_y = -\partial_0 B_x. \qquad (9)$$

The primary wave propagates (in the $y-z$ plane, making an angle $\theta_1$ with the positive $z$-axis) towards the interface ($z = 0$), which separates the two media of dielectric permittivities $\varepsilon_{1,3} = n_{1,3}^2$. In region 2, *which is the thin layer itself*, the field configuration generates an induced current; in this region a term $4\pi j_{2y}/c$ has to be added to the rhs of the first equation of (9). In region 1 we take $B_x$ as a superposition of the *given incoming plane wave pulse* $F$ (of *arbitrary* temporal variation), and an *unknown reflected plane wave* $f_1$. The corresponding electric field components $E_y$ and $E_z$ can be derived from (9),

$$B_{1x} = F - f_1 = F[t - n_1(y \sin \theta_1 - z \cos \theta_1)/c] - f_1[t - n_1(y \sin \theta_1 + z \cos \theta_1)/c],$$
$$E_{1y} = (\cos \theta_1 / n_1)(F + f_1), \quad E_{1z} = (\sin \theta_1 / n_1)(F - f_1). \qquad (10)$$

In region 3 the magnetic induction $B_{3x}$ is represented by the *unknown refracted wave* $g_3$ and, by putting this into (9), the electric field strength can also be obtained,

$$B_{3x} = g_3[t - n_3(y \sin \theta_3 - z \cos \theta_3)/c], \quad E_{3y} = (\cos \theta_3 / n_3)g_3, \quad E_{3z} = (\sin \theta_3 / n_3)g_3. \qquad (11)$$

The relevant boundary conditions for the tangential components read

$$[E_{1y} - E_{3y}]_{z=0} = 0, \quad [B_{1x} - B_{3x}]_{z=0} = (4\pi/c)K_{2y}. \qquad (12)$$

On the basis of equations (9), (10), (11) and (12), the unknown scattered fields $f_1$ and $g_3$ can be expressed in terms of the by now *unknown induced surface current* $K_{y2}$,

$$f_1(t') = (1/(c_1 + c_3))[(c_3 - c_1)F(t') - c_3(4\pi/c)K_{2y}(t'), \quad c_1 = \cos \theta_1/n_1, \quad c_3 = \cos \theta_3/n_3, \qquad (13)$$

$$c_3 g_3(t') = E_{3y}(t') = (2c_1 c_3/(c_1 + c_3))[F(t') - (2\pi/c)K_{2y}(t')], \quad K_{2y} = e\eta v_y, \qquad (14)$$

where $t' = t - yn_1 \sin \theta_1/c$ denotes the retarded time parameter at the surface. Snell's law of refraction ($n_1 \sin \theta_1 = n_3 \sin \theta_3$) means that this retarded time $t'$ must be equal to $t'' = t - yn_3 \sin \theta_3/c$. Equations (13) and (14) are valid for any (constant) $n_{1,3}$, regardless of the nature of $K_{y2}$. On the other hand, since $n_{1,3}$ in general may depend on the frequencies of the Fourier components of the radiation fields, one should separately discuss in each case, whether to what extent the constancy of $n_{1,3}$ applies. Anyway, if the thin layer is in vacuum (air), as for instance a suspended graphene sheet above a trench, then $n_{1,3} = 1$, and no such problem appears. The unknown surface current $K_{2y} = e\eta v_y$ has been expressed (Varró, 2004, 2007a-b-c and 2012) as the product of the surface charge density $e\eta$ and the velocity $v_y = d\delta_y(t')/dt'$ associated to the local displacement $\delta_y(t')$ of the electrons. The original current term $4\pi j/c$ in the Maxwell equations (from which we obtained $4\pi K/c$) could be phenomenologically interrelated to the electric field strength by the constitutive relation $\mathbf{j} = \sigma(\omega) \cdot \mathbf{E}$, where $\sigma(\omega)$ is the conductivity of the layer at a particular frequency of a stationary field (Nair et al (2008), Cserti (2012)). Anyway, it can be shown that the boundary conditions (12), and solutions (13), (14) are consistent with the discontinuty of the displacement field $\mathbf{D}$. By choosing the 'Ansatz' $K_{2y} = e\eta v_y$, we can derive $[D_{1z} - D_{3z}]_{z=0} = 4\pi e \eta$ from them. This relation, on the other hand, can also be obtained from the Gauss law, $\nabla \cdot \mathbf{D} = 4\pi e \rho$, where $e\rho$ is the volume charge density. The form of the surface current we are using corresponds to a *convective* flow, and we consider $\eta$ as an average density.

It is interesting to note that in our procedure the unknown surface current in (12) plays a role of a sort of 'active boundary', contributing to the matched fields. Of course, the force terms in the equation of motion of the surface charge elements must contain the *total field* (including the by now unknown scattered ones). In this way, the present formalism automatically accounts for a '*collective radiation reaction*'; in fact, we describe the dispersion of the layer, as a whole.

In order to solve the scattering problem we use the relativistic equation of motion (8), derived in section 2, for the massless current elements under the action of the composed fields (10) and (11), which have been expressed in (13) and (14). In the TM configuration the $\propto \mathbf{v} \times \mathbf{B}$ term has zero components in the $x-y$-plane. On the other hand, the total electric field has an extra term stemming from the surface current in (13) and (14), which is proportional with the velocity component $\mathrm{v}_y$. Since in the present case $\pi_x = $ constant, the derived equation of motion contains $\mathrm{v}_y$, as the only unknown function,

$$\frac{d\pi_y(t')}{dt'} = \frac{2c_1c_3}{c_1+c_3}\left[eF(t') - \frac{2\pi e^2}{c}\eta \mathrm{v}_y(t')\right], \qquad \mathrm{v}_y = \frac{\mathrm{v}_F}{|\boldsymbol{\pi}|}\pi_y, \tag{15}$$

where the geometrical factors $c_1$ and $c_3$ have been defined in (13). If one finds the solution of (15), then from (10), (11), (13) and (14) the scattered fields can be obtained. On the rhs of (15) the second term in the bracket represents the *collective radiation back-reaction*, which always represents a *damping term*, regardless of the sign of the charges. We represent the incoming field as

$$F(t') = -\partial^2 Z_y / c^2 \partial t'^2, \quad Z_y = -(c^2 F_0 / \omega_0^2) f(t')\cos(\omega_0 t' + \varphi_0), \quad f(t) = \exp(-t^2/2\tau), \tag{16}$$

where $Z_y$ is the Hertz potential of the incoming laser pulse of central frequency $\omega_0$, field strength $F_0$, carrier-envelope (CE) phase difference $\varphi_0$, and $\tau = \tau_L / 2\sqrt{\log 2}$. In the envelope function $\tau_L$ means the full temporal width at half maximum of the intensity. By introducing the dimensionless variables $\omega_0 t' / 2\pi = t'/T \to t$ and $c\boldsymbol{\pi}/\hbar\omega_0 \to \mathbf{q}$, equation (15) becomes

$$\frac{dq_y(t)}{dt} = \frac{2c_1c_3}{c_1+c_3}\left[\pi R\mu_0\{F(t)/F_0\} - \alpha\beta_F\eta\lambda_0^2 \frac{q_y(t)}{\sqrt{q_x^2 + q_y^2(t)}}\right], \quad \mathbf{q} \equiv \frac{c\boldsymbol{\pi}}{\hbar\omega_0}, \quad t \equiv \frac{\omega_0 t'}{2\pi} = \frac{t'}{T},$$

$$\alpha \equiv \frac{e^2}{\hbar c}, \qquad \beta_F \equiv \frac{\mathrm{v}_F}{c}, \qquad R \equiv \frac{2mc^2}{\hbar\omega_0}, \qquad \mu_0 \equiv \frac{eF_0}{mc\omega_0} = 10^{-9}S^{1/2}/E_{ph}, \qquad \pi R\mu_0 \equiv \bar{\mu}_0. \tag{17}$$

In equation (17) $\alpha \approx 1/137$ is the fine structure constant, $\beta_F \approx 1/300$, $\lambda_0 = 2\pi c/\omega_0$ is the central wavelength of the radiation, and $m$ the usual rest mass of the electron. The parameter $\mu_0$ is the 'dimensionless vector potential', which has earlier been called 'dimensionless intensity parameter'. Its numerical value can be calcuated from the displayed formula, where $S \equiv [I/(W/cm^2)]$ is the intensity, measured in W/cm$^2$, and $E_{ph} \equiv [\hbar\omega_0 / eV]$ is the photon energy, measured in electron Volts. We see that $\bar{\mu}_0$ can be much larger than $\mu_0$, at the same intensity, because $R$ is on the order of $10^6$ in the optical region (i.e. $\hbar\omega_0 = 1eV$). The definition $\mathbf{q} \equiv c\boldsymbol{\pi}/\hbar\omega_0$ corresponds to the scaling of the kinetic momentum by the central photon momentum. We also note that in the geometrical arrangement (p-polarization) under discussion, the $x$−component $q_x$ is a constant of motion.

In the next section we shall present a couple of numerical examples, showing the temporal evolution of the electric field strength of the scattered radiation, and illustrate the "relativisting clipping effect", caused by the interaction with the charged layer.

## 4. Numerical examples illustrating the relativistic clipping effect

The first term in the scattered field $f_1(t)$ given by (13) vanishes if $c_3 = c_1$, which condition is equivalent to $\theta_1 + \theta_3 = \pi/2$. This means that if the angle of incidence is the Brewster–angle, then only the surface current term $K_{2y} = e\eta v_y$ contributes to the reflected field. By leaving out this term, we get back the Brewster phenomenon (the p-polarized component in the reflected radiation gets to zero), which is described by the usual Fresnel formulae (see e.g. Born and Wolf (1959)). We note that this condition is satisfied at *any angles of incidence* if $n_3 = n_1$, in particular, if the layer is in vacuum (air). In these cases

$$f_1^{(Brewster)} = -(2\pi e\eta/c)v_y = -2\pi e\eta \frac{v_F}{c} \cdot \frac{q_y}{|\mathbf{q}|} \quad (c_3 = c_1), \quad f_1^{(max)} = 3016(V/cm)\times 10^{-12}(\eta cm^2). \quad (18)$$

This field has the upper bound $|f_1^{(Brewster)}| \leq 2\pi(v_F/c)e\eta$, which does not depend on the intensity. The numerical value of this upper bound is $f_1^{(max)} = 3016(V/cm)$, if we chose as an illustration the surface density $\eta = 10^{12}cm^{-2}$. It is 'universal', in the sense that it does not depend either on the frequency or on the wavelength. The field strength $3016(V/cm)$ corresponds to the intensity $12064W/cm^2$, which is equivalent to the intensity of a 5711 K temperature black–body radiation, according to the Stefan–Boltzmann law.

In the first numerical example we take $\lambda_0 = 620nm$, i.e. $\hbar\omega_0 = 2eV$, and $\eta = 10^{12}cm^2$, in which case in the differential equation (17) we have the parameter value $\alpha\beta_F\eta\lambda_0^2 = 0.094$. In Figure 1 we present the normalized electric field strenght of an $I = 1MW/cm^2$ intensity incoming field, in which case the driving term in (17) has the numerical value $\bar{\mu}_0 = \pi R\mu_0 = 0.785$, and $\alpha\beta_F\eta\lambda_0^2/\bar{\mu}_0 = 0.12$. At this point we note that $\bar{\mu}_0$ and $\alpha\beta_F\eta\lambda_0^2$ are both proportional to $\lambda_0^2$, thus their relative size does not depend on the wavelentgh of the incoming radiation. The $x$−component of the scaled momentum, $q_x$ is a constant of motion, we have chosen $q_x = 3\times q_y(0)$, and then $q_y(0) = q(0)/2$. We note that we have carried out the calculations for various initial values, but we have not found considerable qualitative changes. In the figures we illustrate the temporal evolution of the electric field strength, as a function of the dimensionless time variable, $(t - ny\sin\theta/c)/T$ drawn on the abcissa. The reflected signal contains only a radiation reaction term which is proportional with the velocity of the surface current (see (18)). Due to the ultrarelativistic kinematics of the electrons, the maximum signal is reached already at the rising part, and this 'saturation' causes a sort of '*relativistic clipping*', because the velocity component cannot increase beyond $v_F$. This results in a nearly *rectangular* temporal shape of $v_y$, even at such a relatively modest intensity as $I$=$1MW/cm^2$.

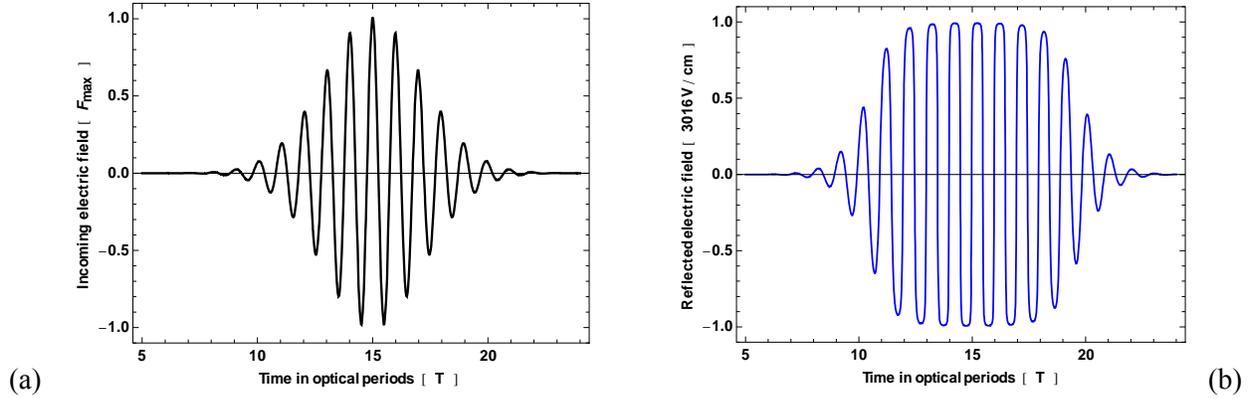

**Figure 1.** Shows the distortion of a 3-cycle p-polarized incoming Gaussian laser pulse (left; Figure 1a) impinging on a graphene layer at Brewster angle (or at *any* angle *when* $n_3=n_1$). We have taken for the CE-phase $\phi_0=0$ (cosine pulse), and $\lambda_0=620nm$, $I=1MW/cm^2$, the effective intensity parameter is $\pi R\mu_0=0.785$. The damping parameter used in the differential equation (17) equals $\alpha\beta_t\eta(\lambda_0)^2=0.094$. The reflected component is nonzero (right; Figure 1b), its existence cannot come out from the usual Fresnel formulae. The units of the electric field on the ordinate in Figure 1b is $3016V/cm$. After saturation, this 'universal', maximum value is reached (for a surface density $\eta=10^{12}cm^{-2}$, according to the definition in (18)). The distorted shape stems from the 'clipping effect', discussed in the main text.

In Fig 1b the nearly rectangular structure resembles to the orthogonal sytem of functions $\{r_k(x)\}$ in the unit interval introduced by Rademacher (1922),

$$r_0(x)=1, \quad r_k(x)=sign(\sin 2^k\pi x) \quad (k=1, \ 2, \ ...). \tag{19}$$

The spectrum corresponding to the Rademacher functions is proportional to $1/(2n+1)^2$ or $1/(2n+2)^2$, depending on parity, where $n$ is the harmonic index. We note that this sytem (or rather, the derived complete system of the so-called Walsh functions) is used in signal analysis and synthesis in the *microwave regime*. Here we found a possible physical representatives of them in the *optical or terahetrz regime*. This is better illustrated in Figures 2 and 3.

In Figure 2 we show the temporal behaviour in case of a larger intensity, $I=100MW/cm^2$: $\bar\mu_0=7.85$.

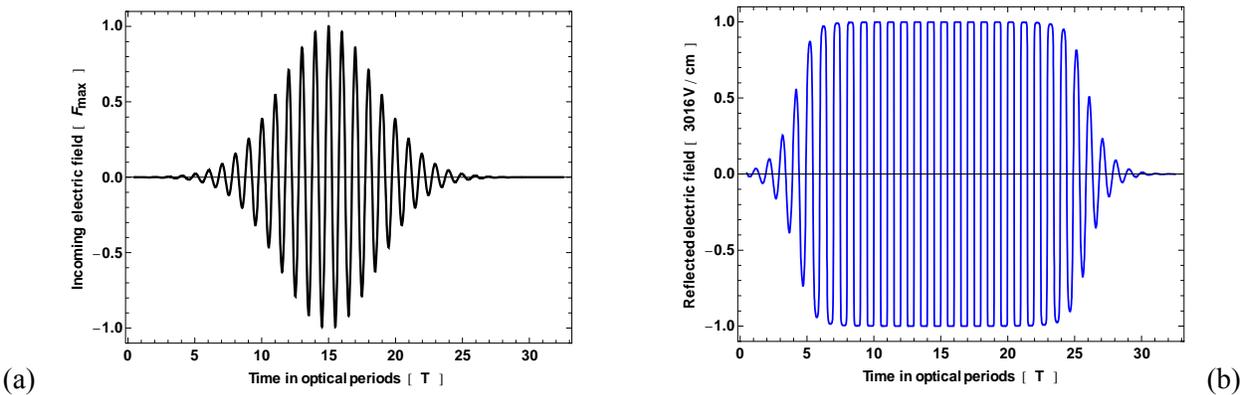

**Figure 2.** Illustrates the distortion of a 5-cycle p-polarized incoming Gaussian laser pulse (left; Figure 2a) impinging on a graphene layer at Brewster angle (or at *any* angle *when* $n_3=n_1$). We have taken $I=100MW/cm^2$, then the effective intensity parameter is $\pi R\mu_0=7.85$. The other parameters are the same as in Figure 1. The reflected component is shown on the right (Figure 2b), whose nearly rectangular shape again stems from the 'clipping effect'. It should be noticed that the reflected signal reaches its maximum already in the rising part of the incoming pulse.

Our last example is a the scattering of a sub–cycle pulse (~0.8–cycle) similar to that produced recently by Wirth *et al* (2011). The peak field strength shown in their Figure 3B was $4 \times 10^{7} V/cm$, which corresponds to the intensity $2.122 \times 10^{12} W/cm^{2}$, according to the conversion $F_{0}/(V/cm) = 27.46 \times \sqrt{I_{0}/(W/cm^{2})}$. Instead of this large intensity, we take $I = 1 GW/cm^{2}$ only, and then the effective intensity parameter of the incoming pulse becomes $\bar{\mu}_{0} = 24.82$. In Figure 3 we show the temporal behaviour of such a pulse and the response of the layer.

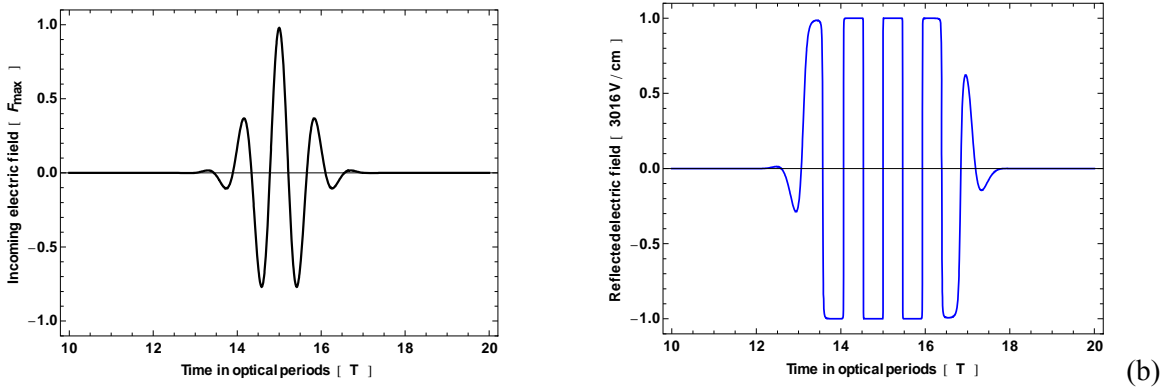

(a)                                                  (b)

**Figure 3.** Illustrates the distortion of a sub-cycle (0.8-cycle) p-polarized incoming Gaussian laser pulse (left; Figure 3a) impinging on a graphene layer at Brewster angle (or at *any* angle *when* $n_{3}=n_{1}$). We have taken $I$=1GW/cm², then the effective intensity parameter is $\pi R \mu_{0}$=24.82. The other parameters are the same as in Figures 1 and 2. The reflected component is shown on the right (Figure 3b), its nearly rectangular shape again stems from the 'clipping effect'. It should be noticed here, too, that the reflected signal reaches its maximum already in the rising part of the incoming pulse. Its shape is essentially the same as a Rademacher function on the interval (-3T/2,+3T/2).

As one can see in figures 1, 2 and 3, according to our description above, the optical response of graphene really causes high nonlinearities. On the other hand, due to this same violent response of the relativistic charges, a very stong radiation damping develops. The interplay of these two effects, through the ultrarelativistic kinematics, results in an almost universal rectangular temporal evolution of the reflected signal. We note that this structure follows the shifts of the CE–phase, as has recently been discussed by us (Varró, 2012). We also note that, that the tails of the reflected signals can be considerably longer than it is shown for instance in Fig 3b. Such wake–fields (Varró, 2007a-b-c) may form slowly decreasing rectified quasi–static fields.

## 5. Summary

We have presented our recent theoretical results on the reflection and transmisson of few-cycle laser pulses on a very thin conducting layer, which has been represented by a classical convective surface current density of the massless relativistic charges. We have shown that the pulses can undergo considerable distortions, even at relatively modest laser intensities. They are deformed to rectangular trains, which may be considered as approximate physical realizations of Rademacher functions in the optical or terahertz regime. In Section 2 we derived the gauge–invariant equation of motion of massless charges. In Section 3 the matching equations for the electromagnetic fields, and the radiation reaction term in the equation of motion have been determined. In Section 4 the nature of the 'relativistic clipping effect' has been discussed, with inclusion of some illustrative figures. The numerical values of the input parameters have been chosen to correspond to the massless electrons and holes of a graphene monolayer.

We would like to emphasize that, it has long been known that the velocity function of the 'ordinary' massive electrons can also have an essentially rectangular shape at ultrarelativistic ($I >> 10^{18} W/cm^2$) intensities (see e.g. Varró (2010)). Besides, even at relatively moderate intensities, the nonlinearity parameter can also be so large that a similar distortion happens in the induced current in crystals (see the expression for the current density and Figure 5b in the paper by Ghimire *et al* (2011)). However, in each cases the radiated field is proportional with the acceleration, not with the velocity, as in the present analysis. Roughly speaking, if, for the former case we took the derivative of the Rademacher functions (with half-cycle constancy regions of the velocity function), then we would receive very short (attosecond) peaks at the (approximate) discontinuty points.

The relativistic clipping effect described above can in principle manifest itself at any frequencies; in the terahertz, optical, or even higher frequencies, opening a wide range of potential applications. However, for the large ferqency responses the model of massless charges may not be relevant. One should also keep in mind that for higher frequencies or/and larger intensities the linear or/and nonlinear photoelectic effect may be an important competing process in real systems, which has been completely left out of the above considerations. Besides, though we think that from the present classical description the essentials of the process can be understood, a quantum calculation is also desirable.

## Acknowledgments

This work has been supported by the Hungarian National Scientific Research Foundation OTKA, Grant Nos. K73728 and K 104260. We thank the valuable discussions with Prof. J. Cserti and dr. J. K. Asbóth on the capabilities and limitations of our model of massless charges. Partial support by the National Development Agency, Grant No. ELI_ 09-1-2010-0010, Helios Project is also acknowledged.

## References

Abergel D S L and Fal'ko V I 2007 Optical and magneto-optical far-infrared properties of bilayer graphene 2007 *Phys. Rev. B* 75 155430

Asbóth J K 2012 Symmetries, topological phases, and bound states in the on-dimensional quantum walk *Phys. Rev. B* **86** 195414

Blake P, Hill E W, Castro Neto A H, Novoselov K S, Jiang D, Yang R, Booth T J and Geim A K 2007 Making graphene visible *Appl. Phys. Lett.* **91** 063124

Born M and Wolf E 1959 *Principles of Optics* (London: Pergamon, 1959)

Bostwick A, Ohta T, Seyller T, Horn K and Rotenberg E 2007 Quasiparticle dynamics in graphene *Nat. Phys.* **3** 36-40

Castro Neto A H, Guinea F, Peres N M R, Novoselov K S and Geim A K 2009 The electronic properties of graphene *Rev. Mod. Phys.* **81**, 109-162

Cserti J and Dávid Gy 2010 Relation between Zitterbewegung and the charge conductivity, Berry curvature, and the Chern number of multiband systems *Phys. Rev. B* **82** 201405(R)

Cserti J 2012 Private communication, Unpublished note.

Cruz y Cruz S, Negro J and Nieto L M 2008 On position dependent mass harmonic oscillator *J. Phys: Conference Series* **128** 012053 (V International Symposium on Quantum Theory and Symmetries)

Fedorov M V 1997 *Atomic and Free Electrons in a Strong Laser Field* (World Scientific, Singapore)

Földi P and Benedict M G 2012 Ultrashort pulse induced currents in solids: Theoretical approaches *Light at Extreme Intensities 2011* AIP Conf. Proc. **1462** 96-99

Ghimire S, DiChiara A D, Sistrunk E, Agostini P, DiMauro L F and Reis D A 2011 Observation of high-order harmonic generation in a bulk crystal *Nature Physics* 7 138-141

Hasan M Z and Kane C L 2010 Topological insulators *arXiv*:1002.3895v2 [cond-mat.mes-hall]

Varró S,  Generation of rectangular optical waves by relativistic clipping. [21th International Laser Physics Workshop (LPHYS'12) July 23-27, 2012, Calgary, Canada (2.7.5).]

Korbman M, Kruchinin S Yu and Yakovlev V S 2012 Quantum beats in the polarization response of a dielectric to intense few–cycle laser pulses *arXiv*:1210.2238v2 [physics.atom-ph]

Krausz F and Ivanov M 2009 Attosecond physics *Rev. Mod. Phys.* **81** 163-234

Milošević D B, Paulus G G, Bauer D and Becker W 2006 Above–threshold ionization by few–cycle pulses *J. Phys. B: At. Mol. Opt. Phys.* **39** R203-R262

Nagele S, Pazourek R, Feist J, Doblhoff-Dier K, Lemell C, Tőkési K and Burgdörfer J 2011 Time-resolved photoemission by attosecond streaking: extraction of time information *J. Phys. B: At. Mol. Opt. Phys.* **44** 081001

Nair R R, Blake P, Grigorenko, Novoselov K S, Booth T J, Stauber T, Peres N M R and Geim A K 2008 Fine structure constant defines visual transparency of graphene *Science* **320** No5881 p1308

Neppl S, Ernstorfer R, Cavalieri A L, Menzel D, Barth J V, Krausz F, Kienberger R and Feulner P 2012 Attosecond time-resolved photoemission from core and valence states of magnesium *Phys. Rev. Lett.* **109** 087401

Novoselov K S, Geim A K, Morozov S V Morozov S V, Jiang D, Katsnelson M I, Grogorieva I V, Dubonos S V and Firsov A A 2005 Two-dimensional gas of massless Dirac fermions in graphene *Nature* **438**, 197-200

Rademacher H 1922 Einige Sätze über Reihen von allgemeinen Orthogonalfunktionen *Math. Ann.* **87** 112-138

Schultze M, Fieß M, Karpowicz N,  Gagnon J, Korbman M, Hofstetter M, Neppl S, Cavalieri A L, Komninos Y, Mercouris Th, Nicolaides C A, Pazourek R, Nagele S, Feist J, Burgdörfer J, Azzeer A M, Ernstorfer R, Kienberger R, Kleineberg U, Goulielmakis E, Krausz F, and Yakovlev V S 2010 Delay in photoemission *Science* **328** (5986) 1658–1662

Uiberacker M, Uphues Th, Schultze M, Verhoef A J, Yakovlev V S, Kling M F, Rauschenberger J, Kabachnik N M, Schröder H, Lezius M, Kompa K L, Muller H-G, Vrakking M J J, Hendel S, Kleineberg U, Heinzmann U, Drescher M and Krausz F 2007 Attosecond real-time observation of electron tunnelling in atoms *Nature* **446** (7136) 627–632

Varró S 2004 Scattering of a few-cycle laser pulse on a thin metal layer: the effect of the carrier–envelope phase difference *Las. Phys. Lett.* **1** 42-45

Varró S 2007a Reflection of a few–cycle laser pulse on a metal nano–layer: generation of phase dependent wake–fields  *Las. Phys. Lett.*  **4** 138-144

Varró S 2007b Scattering of a few–cycle laser pulse by a plasma layer: the role of the carrier–envelope phase difference at relativistic intensities *Las. Phys. Lett.* **4** 218-225

Varró S  2007c  Linear and nonlinear absolute phase effects in interactions of ulrashort laser pulses with a metal nano-layer or with a thin plasma layer *Laser and Particle Beams* **25** 379-390

Varró S 2010 Intensity effects and absolute phase effects in nonlinear laser–matter interactions  *Laser Pulse Phenomena and Applications* (Edited by F. J. Duarte; InTech, Rijeka) Chapter 12, pp 243-266

Varró S 2012 Graphene-based carrier-envelope phase difference meter  *Light at Extreme Intensities 2011* AIP Conf. Proc. **1462** 128-131

Wirth A, Hassan M Th, Grguras I, Gagnon J, Moulet A, Lou T T, Pabst S, Santra R, Alahmed Z A, Azzeer A M, Yakovlev V S, Pervak V, Krausz F and Goulielmakis E 2011 Synthesized light transients *Science* **334** 195-200

Zhang C-H and Thumm U 2011 Probing dielectric-response effects with attosecond time-resolved streaked photoelectron spectroscopy of metal surfaces *Phys. Rev. A* **84** 063403